# Nanomechanical characterization of quantum interference in a topological insulator nanowire

**Authors:** Minjin Kim[1], Jihwan Kim[2], Yasen Hou[3], Dong Yu[3], Yong-Joo Doh[4], Bongsoo Kim[1], Kun Woo Kim[5*] and Junho Suh[2*]

**Affiliations:**

[1]Department of Chemistry, Korea Advanced Institute of Science and Technology, Korea

[2]Quantum Technology Institute, Korea Research Institute of Standards and Science, Korea

[3]Department of Physics, University of California at Davis, U.S.A.

[4]Department of Physics and Photon Science, Gwangju Institute of Science and Technology, Korea

[5]Center for Theoretical Physics of Complex Systems, Institute for Basic Science (IBS), Daejeon, Korea

*Correspondence to: kkimx4@ibs.re.kr, junho.suh@kriss.re.kr

**The discovery of two-dimensional (2D) gapless Dirac fermions in graphene[1] and topological insulators (TI)[2] has sparked extensive ongoing research toward applications of their unique electronic properties[3–6]. The gapless surface states in three-dimensional (3D) insulators indicate a distinct topological phase of matter with a non-trivial $Z_2$ invariant[6,7] that can be verified by angle-resolved photoemission spectroscopy[2,8,9] or magnetoresistance quantum oscillation[10,11]. In TI nanowires, the gapless surface states exhibit Aharonov–Bohm (AB) oscillations in conductance[12–14], with this quantum interference effect accompanying a change in the number of transverse one-dimensional (1D) modes in transport[15]. Thus, while the density of states (DOS) of such nanowires is expected to show such AB oscillation, this effect has yet to be observed. Here, we adopt nanomechanical measurements[16–19] that reveal AB oscillations in the DOS of a topological insulator. The TI nanowire under study is an electromechanical resonator embedded in an electrical circuit, and quantum capacitance effects from DOS oscillation modulate the circuit capacitance thereby altering the spring constant to generate mechanical resonant frequency shifts. Detection of the quantum capacitance effects from surface-state DOS is facilitated by the small effective capacitances[20] and high quality factors[21,22] of nanomechanical resonators, and as such the present technique could be extended to study diverse quantum materials at nanoscale.**

We design our device to measure the fundamental mode frequency of flexural vibration in a suspended TI nanowire fabricated with single-crystalline $Bi_2Se_3$ (Fig. 1a). The mechanical motion of the nanowire changes the electrostatic energy stored in a circuit, from which shifts in resonant frequency can be computed. In the circuit in Fig. 1b, as DC gate voltage $V_g$ is introduced, charges are induced to compensate for the potential difference between the gate and nanowire. At the same time, nanowire chemical potential $\mu$ is changed by induced charge $Q$ as

$$V_g = \frac{Q}{C_G} + \frac{\mu - \mu_0}{e}, \quad (1)$$

where $C_G$ is the geometric capacitance between the nanowire and gate electrode, $\mu_0$ is the intrinsic chemical potential of a nanowire with $V_g = 0$, and $e$ is the elementary charge. In addition to $V_g$, a radio-frequency (RF) voltage $V_{RF}$ is applied to induce nanowire vibration, a motion that modifies $C_G$ followed by the modulation of $Q$ and $\mu$ according to equation (1). The amplitude of charge modulation measured as $V_{OUT}$ is maximized when $V_{RF}$ meets the mechanical resonant frequency[20]. In this way, we trace the mechanical resonant frequency by sweeping the gate voltage $V_g$ (Fig. 1c) or magnetic field $B$. Note that the mechanical vibration (~100 MHz) is much slower than the typical time scales of electron dynamics considered here (relaxation time ~1 ps, see Supplementary Information section S6), and we assume the electrostatic equation to hold throughout.

If the density of states is infinite as in metals, chemical potential remains at $\mu_0$ regardless of $Q$. With the finite DOS of TI nanowires, however, change in gate voltage is divided into the potential difference between gate and wire and the change in chemical potential, as

$$\delta V_g = \frac{\delta Q}{C_G} + \frac{\delta \mu}{e} = \delta Q \left( \frac{1}{C_G} + \frac{1}{e^2 L \nu} \right), \quad (2)$$

where $\nu = \delta(Q/Le)/\delta\mu$ is the number of electronic states per unit length per unit energy, and $L$ is wire length. From this relation, we introduce quantum capacitance to account for the potential difference made internally by the induced charge[16], $C_Q = e^2 L \nu$. In our experiments, we estimate that quantum capacitance dominates geometric capacitance ($C_Q/C_G \sim 10^3$, see Supplementary

Information section S3 and S4), and thus the first term in equation (2) accounts for the majority of gate voltage change. However, AB oscillation in the mechanical resonant frequency turns out to originate mostly from the second term related to DOS in our experiment.

When a magnetic flux ($\Phi$) is applied along the nanowire axis, 1D sub-band energy is described by[12,15]

$$\varepsilon(n, k, \Phi) = \pm \hbar v_F \sqrt{k^2 + \frac{(n + 1/2 - \Phi/\Phi_0)^2}{R^2}}, \quad (3)$$

where $n$ is the sub-band index, $k$ is a 1D momentum vector along the nanowire direction, $h$ is Planck's constant, $v_F$ is the Fermi velocity[23] ($\approx 5\times 10^5$ m/s for $Bi_2Se_3$), $R$ is nanowire radius, and $\Phi_0$ is the flux quantum ($= h/e$). In cylindrical geometry, a gapless Dirac fermion picks up the Berry phase of $\pi$ by encircling the perimeter, and a gapless 1D mode appears when an external magnetic field supplies an additional AB phase $\pi$. As an example, we draw the band dispersion of 1D modes for $\Phi = 0$ and $\Phi = \Phi_0/2$ in Fig. 1d; appearance of the 1D gapless mode is followed by an energy shift of neighboring transverse modes, thus also changing conductance $G$ at a finite energy (Fig. 1e, top).

The appearance of the gapless mode also affects the DOS (Fig. 1e, bottom), with related quantum capacitance effects resulting in mechanical resonant frequency shifts that can be obtained by taking the variation of electrostatic energy (Fig. 1f). Induced charge $Q$ is computed by integrating the DOS from the initial chemical potential $\mu_0$ to $\mu$, as $Q = Le \int_{\mu_0}^{\mu} \nu(E) dE$, where the sign of $Q$ is negative for $\mu < \mu_0$ (or $V_g < 0$). With mechanical vibration, change in electrostatic energy for a fixed gate voltage reads

$$\delta U_{\text{ec}} = \delta\left(\frac{Q^2}{2C_G}\right) - V_g \delta Q + \delta\left(Q\frac{\mu - \mu_0}{e}\right), \quad (4)$$

where the first, second, and third terms are from the energy stored in the geometric capacitance, work done to the $V_g$ source, and quantum capacitance charging energy, respectively. By taking the second derivative of $U_{\text{ec}}$ with respect to nanowire displacement $x$, we obtain the change in spring constant due to electrostatic energy by

$$k - k_0 \approx -\frac{1}{2}\left(\frac{\ddot{C}_G}{C_G^2}\right)Q^2 + \frac{e}{2}\left(\frac{\dot{C}_G}{C_G}\right)^2 \frac{\partial}{\partial \mu}\left(\frac{1}{C_Q^2}\right)Q^3 = k_I + k_{II}, \quad (5)$$

where the dots indicate derivatives with respect to $x$, $k_0$ is the bare effective spring constant of mechanical resonator, and $C_G \ll C_Q$ is assumed (Supplementary Information section S5). We are mainly interested in the modulation of resonant frequency with respect to magnetic field: $\Delta k_{I,II} = k_{I,II} - \langle k_{I,II} \rangle_B$ where $\langle \ldots \rangle_B$ refers to the averaged value over magnetic field at a given $V_g$. Mechanical resonant frequency shifts corresponding to $\Delta k_{I,II}$ are given by $\Delta f_{I,II} = (\Delta k_{I,II}/2\langle k \rangle_B) \cdot \langle f \rangle_B$. In equation (5), $k_I$ is easily understood as the gate capacitance effect[24] with a correction of induced charge due to the magnetic flux and chemical potential. Otherwise, the second term $k_{II}$ indicates a novel effect particularly important to our 1D TI system, as it contains a derivative of the DOS (or quantum capacitance) with respect to chemical potential as well as the third-order power of induced charge. Considering that $Q$ is negative (i.e., $V_g < 0$, where our extensive measurements are conducted), the DOS derivative in $\Delta k_{II}$ makes an important contribution (Fig. 1f): as the chemical potential approaches the Dirac point where the DOS profile gets sharper with less scattering, $\Delta f_{II}$ grows significantly faster than $\Delta f_I$. Accordingly, at higher energies with larger DOS, $\Delta k_I$ dominates.

We first characterize the AB conductance oscillation[9] in our TI nanowire by measuring conductance at various magnetic fluxes and gate voltages (Fig. 2a). The magneto-conductance $\Delta G$ oscillates with a period of $\Delta B = 0.4$ T, as its fast Fourier transform (FFT) confirms (Fig. 2a, inset). The effective cross-section, $\Phi_0/\Delta B = 1.04 \times 10^4$ nm$^2$, is consistent with nanowire dimensions (Supplementary Information section S1). Conductance modulation occurs with gate voltages as well, and correlation between two conductance traces at $\Phi = 1.5\ \Phi_0$ and $2\ \Phi_0$ is out of phase, as expected from TI surface states (Fig. 2b)[12–14]. The periodicity estimated from the FFT (Fig. 2b, inset) is $\Delta V_g \approx 5.7$ V, which corresponds to the level spacing between neighboring transverse modes $\Delta = \frac{\hbar v_F}{R} \cong 4.7$ meV. Since the DOS changes over energy, gate voltage is not linearly proportional to chemical potential change, with approximately $\mu - \mu_0 \sim V_g^2$ for 2D Dirac dispersion. Thus, change in total conductance is also a quadratic function of gate voltage, as the chemical potential of the TI nanowire approaches the Dirac point (Fig. 2c, inset). In Fig. 2c, total conductance difference between $V_g = -32$ V and $-20$ V, corresponding to a chemical potential difference of $2\ \Delta$, is approximately 2.5 $(e^2/h)$. Thus, we estimate that the chemical potential is near $\mu \approx 24\ \Delta$ when total conductance $G \approx 30\ (e^2/h)$, indicating that our experiments are executed far away from the conduction band edge ($\sim 50\ \Delta$)[9,23], and so bulk-state contributions are not considered in our analysis.

Our TI nanowire device setup allows for the simultaneous measurements of electrical conductance and mechanical resonant frequency, with the observation of strong correlation between the two independent signals confirming their common origin: AB oscillation. For two representative magnetic fluxes $\Phi = 1.5\ \Phi_0$ and $2\ \Phi_0$, conductance modulation $\Delta G$ and resonant frequency shift $\Delta f_0$ with respect to gate voltage are shown in Fig. 3a–b. We note that the resonant frequency shift is close to being out of phase with conductance modulation. The observed correlation is successfully reconstructed by a quasi-1D wire model of a gapless Dirac fermion (Fig. 3c–f). To compute magneto-conductance, we employ the Landauer–Buttiker formalism for a 3D TI lattice model maintaining the aspect ratio of our nanowire[25,26]. We simulate conductance at the chemical potential in our device by introducing a disorder strength that yields a scattering time close to the one in the nanowire (Supplementary Information section S6). In this way, the observed AB oscillation of conductance is reproduced reasonably well (Fig. 3d–f). For the modulation of mechanical resonant frequency shift, we employ the quasi-1D model of a Dirac fermion with complete eigenenergy information for a given magnetic flux from equation (3). Using a scattering time consistent with the disorder strength used in the magneto-conductance calculation, we first obtain the DOS from the Green's function and then compute the shift of resonant frequency according to equation (5). The calculated frequency modulation successfully explains the experimental data in the range of chemical potential (Fig. 3c, e, f). Most importantly, the out-of-phase relation between conductance and mechanical resonant frequency shift in Fig. 3a–b is clearly reproduced in our model calculations in Fig. 3e–f, confirming the nature of the TI surface states.

We provide a more comprehensive experimental and numerical presentation of the mechanical resonant frequency shift in Fig. 4. The AB oscillations of resonant frequency at different gate voltages are shown in Fig. 4a; as the chemical potential approaches the Dirac point from $V_g = -18.8$ V to $-31.2$ V, an increase in oscillation amplitude is noticeable, as expected from Fig. 1f. The FFT (Fig. 4a, inset) clearly shows that the period of resonant frequency oscillation equals that of conductance, $\Delta B = 0.4$ T. A phase alternation of resonant frequency oscillation takes place with gate voltages as well. The mechanical resonant frequency shift versus $V_g$ curves (Fig. 4c) for integer and half-integer flux quanta also exhibit an out-of-phase relation with each other, and their oscillation period $\Delta V_g \approx 5.7$ V (Fig. 4c, inset) agrees well with conductance oscillation data. We numerically compute the shift of resonant frequency for an extended energy window in Fig. 4b, and observe a qualitative change in the oscillation pattern from a checkerboard-like shape for $\mu > 25\,\Delta$ where $\Delta k_\mathrm{I}$ in equation (5) is larger, to a diamond-like shape for $\mu < 25\,\Delta$ where $\Delta k_\mathrm{II}$ begins to dominate. This crossover must take place at a certain energy in the measurement of 2D Dirac fermions as the chemical potential approaches the Dirac point, where the quantum capacitance holds the majority of potential difference (equation (2)). Not only does the overall shape of the pattern change, but the relative correlation with respect to conductance modulation reverses; for example, note the location of peaks and dips at $\Phi = 0$ along the energy. From total conductance measurements, we estimate the chemical potential to be approximately $24\,\Delta$, which is near the crossover energy. With information that the observed conductance and frequency shift modulations are out of phase with each other (Fig. 3), the chemical potential range in our measurement narrows down to $\mu \cong 23\,\Delta$ to $25\,\Delta$. Further comparisons between experiment and theory along magnetic flux at different gate voltages in Fig. 4e–h confirm this analysis.

The characterization of various topological phases of matter is at the cutting edge of experimental condensed matter physics. Our work presents a novel characterization of a topological phase in $Bi_2Se_3$ nanowire via nanomechanical resonance measurement, which shows more pronounced shift as the chemical potential approaches the Dirac point. This result not only suggests that our nanomechanical detection scheme would be generally applicable to a variety of materials with Dirac electronic structures, but it further provides the novel physics of mechanical motion of a nanostructure combined with non-trivial electronic states.

**Methods**

**Device fabrication.** Single-crystalline $Bi_2Se_3$ nanowires are synthesized by chemical vapor deposition using Au nanoparticles as a catalyst. A single $Bi_2Se_3$ nanowire is transferred with a nano-manipulator to a $SiO_2$/Si substrate with a trench having a bottom-gate electrode. For trench fabrication, poly(methyl methacrylate) (PMMA) resist is spin coated onto the $SiO_2$/Si substrate and the trench area is patterned using electron beam lithography followed by etching the 500 nm-thick-$SiO_2$ layer with buffered oxide etchant (BOE). The bottom-gate electrode is deposited at the bottom of the trench with electron beam evaporation (5 nm Ti, 45 nm Au). After transferring the nanowire, a PMMA resist is spin coated onto the nanowire and baked at 180 °C for 2 min. Selected areas around the nanowire are patterned using electron beam lithography and 20 nm/200 nm Ti/Au electrodes are deposited via AC-sputtering. For metallic contacts to the nanowire, PMMA residue and the native oxide layer on the nanowire surface is removed using a plasma asher and immersing in BOE for 10 s before sputtering. A gate electrode is located approximately $d \sim$ 170 nm below the suspended nanowire, and source and drain contacts are provided to compose a field-effect transistor geometry. The suspended nanowire has dimensions of width 105 nm, thickness 116 nm, and length 1.5 μm.

**Measurements.** The device is measured in a $^3$He/$^4$He dilution refrigerator, and typical measurements are performed at 50 mK. The gate electrode is biased with DC voltage to change the chemical potential of the nanowire, with a sweeping magnetic field along the nanowire direction to modulate the flux through the wire cross-section. Electrical properties of the nanowire are examined by monitoring electrical conductance as measured by comparing near-DC (~17 Hz) source-drain current and voltage across the nanowire. The suspended Bi$_2$Se$_3$ nanowire behaves as a mechanical resonator in response to external force[27]. To detect its fundamental mode, RF voltage is applied to the gate in addition to DC, and the resulting RF current due to gate-capacitance modulation is monitored (Fig. 1b)[18,20]. An example of measured amplitude near one mechanical resonance is plotted in the inset of Fig. 1c. The typical mechanical resonant frequency is about 115 MHz, and the quality factor is about $1.2 \times 10^4$.

**Quasi-1D mode DOS.** The density of states is computed by taking the imaginary part of the retarded Green's function,

$$\nu(E) = -\frac{1}{\pi} \text{Tr}\, Im\left[\frac{1}{E + i\eta - H}\right],$$

$$= \sum_{l,n=-\infty}^{\infty} -\frac{1}{\pi} Im\left[\frac{1}{E + i\eta - \varepsilon(n, k = \frac{2\pi}{L}l, \Phi)}\right],$$

where the dispersion relation $\varepsilon(n, k, \Phi)$ in equation (3) is used. Index $n$ indicates the transverse modes, and $l$ indicates the eigenmodes within the same transverse mode. $\eta$ is a measure of energy broadening caused by elastic scattering from impurities and inelastic scattering from phonons and electron–electron interaction. Thus, it is not straightforward to evaluate scattering time a priori, as only limited microscopic information on the TI nanowire is available. We find that an energy broadening of $\eta = 0.2\, \Delta$, where $\Delta$ is the level spacing between neighboring transverse modes, explains our experimental data well, with a different choice of $\eta$ not affecting the qualitative behavior. Strictly speaking, the energy-broadening $\eta$ is an energy-dependent quantity as more scattering channels are available at higher chemical potential. Nevertheless, since our measurement is carried out within a small energy window (~ 2 $\Delta$), we ignore changes in $\eta$. Instead, for the calculation of resonant frequency shift we made an average over $\mu_0$ as it pertains to larger energy uncertainty. For illustrative purposes, we set $2\pi R/L \ll 1$ and $\eta = 0.05\,\Delta$ in Fig. 1e–f, while we set $2\pi R/L = 1/3$ in Fig. 3c and Fig. 4b for a realistic description of the experiment.

**Lattice model conductance.** Landauer–Buttiker formalism[25] is employed to compute conductance in the quasi-1D modes on the surface of a 3D lattice model for TIs. According to Fisher and Lee[28], DC conductance $G$ of a finite system with static disorder is related to its transmission matrix $t$ by the following relation:

$$G = \frac{e^2}{h} Tr(t^\dagger t),$$

which is the sum of transmission eigenvalues. This expression is valid for any number of scattering channels. The transmission coefficient is then computed from the Green's function in real space representation by taking an element connecting the location of the left lead to that of the right lead. For a memory efficient recursive Green's function method, see the work by Conan[26].

The lattice Hamiltonian[29] employed is

$$H = -t \sum_{n,j=1,2,3} \left( \psi_{n+\hat{e}_j}^\dagger \frac{\Gamma^1 - i\Gamma^{j+1}}{2} \psi_n + \text{h.c.} \right) + \sum_n \psi_n^\dagger (m\Gamma^1) \psi_n + \sum_{n \in \text{surface}} \psi_n^\dagger (V_n \Gamma^0) \psi_n,$$

where the gamma matrices are $\Gamma^{(1,2,3,4)} = (\mathbb{I} \otimes s_z, -\sigma_y \otimes s_x, \sigma_x \otimes s_x, -\mathbb{I} \otimes s_u)$, where $s$ and $\sigma$ are Pauli matrices referring to orbital and spin space, respectively, $\Gamma^0$ is the identity matrix, and $t = 1$ and $m = 1.8$ are used. The first term constitutes nearest-neighbor hopping in the lattice, and the second term provides a constant mass. The third term is onsite random potential with uniform distribution, $V_n \in [-W, W]$. A disorder strength of $W = 4.5\,\Delta$ is used for Fig. 3d (see Supplementary Information section S6 for conductance maps at other disorder strengths, $W = 1.5\,\Delta$, $3.0\,\Delta$, and $6.0\,\Delta$). The open boundary condition is introduced in x- and z-directions with lattice size $N_x = N_z = 8$, and the direction of current $\hat{y}$ is chosen without the loss of generality.

Impurities are introduced only on the open surface to minimize coupling to bulk modes, as our interest is the transport in quasi-1D surface modes in the presence of disorder. To match the ratio between the circumference and length of the TI nanowire in experiment, we set the length of the disordered region $N_y = 96$ to be connected to two semi-infinite leads. Magnetic flux $\Phi$ threading through the nanowire is added in the lattice model by replacement hopping along the x-direction as

$$\psi^\dagger_{n_x+\hat{e}_x,n_y,n_z}(H_x)\psi_{n_x,n_y,n_z} \rightarrow \psi^\dagger_{n_x+\hat{e}_x,n_y,n_z}\left(H_x e^{\frac{i2\pi n_z \Phi}{N_x N_z \Phi_0}}\right)\psi_{n_x,n_y,n_z},$$

which introduces a uniform magnetic field. Because boundary modes have finite penetration depths into the bulk, AB oscillation periodicity is larger than $\Phi = \Phi_0$, while periodicity is also dependent on energy as eigenmodes near the bulk band are more delocalized than on the open surface.

**Acknowledgements**

This work was supported by the Basic Science Research Program through the NRF funded by the Ministry of Science and ICT (2016R1C1B2014713, 2016R1A5A1008184), Korea Research Institute of Standards and Science (KRISS-2018-GP2018-0017), and Institute for Basic Science (IBS-R024-D1).

**Author contributions**

M.K. and J.K. fabricated the devices and performed the experiments. Y.H., D.Y., Y.-J.D., and B.K. helped in sample fabrication and characterization. K.W.K. performed theoretical modeling. J.S.

designed the project. M.K., K.W.K., and J.S. analysed the data and prepared the manuscript. All the authors contributed to discussion and manuscript preparation.

**Competing interests**

The authors declare no competing financial interests.

**Figures**

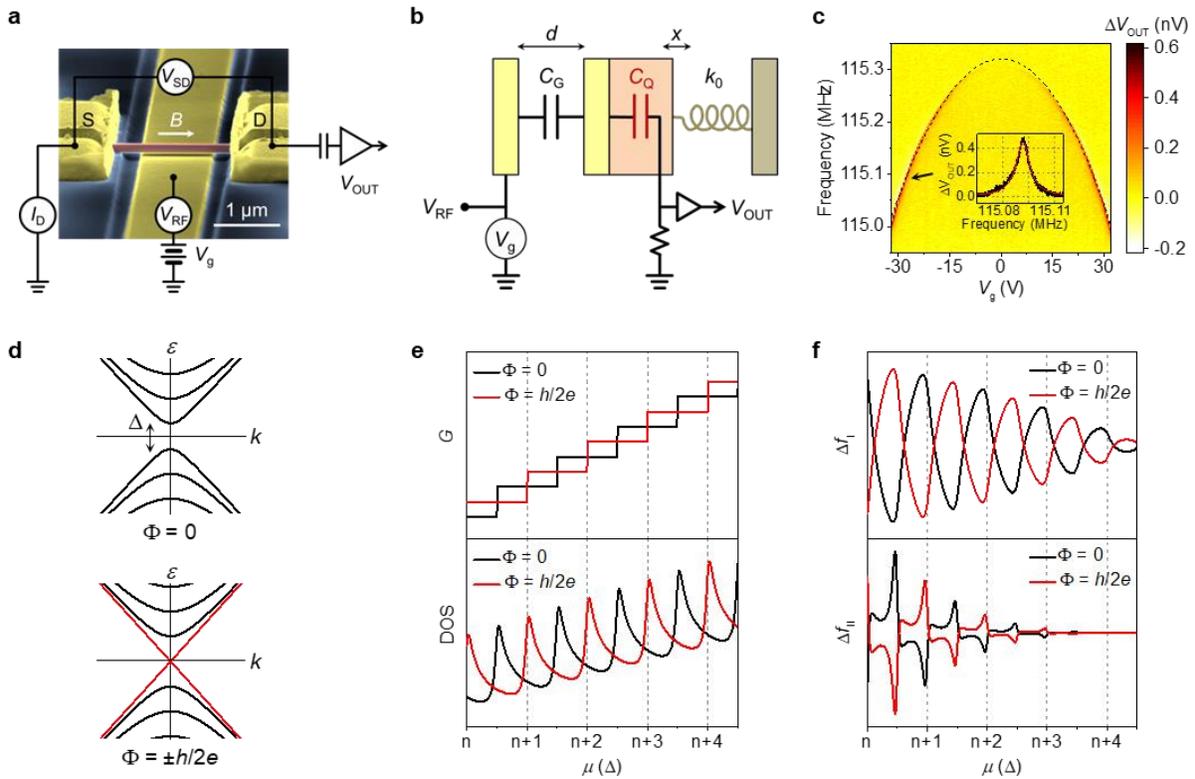

**Figure 1| Device configuration and expected Aharonov–Bohm (AB) oscillations in conductance and mechanical resonant frequency. a**, Scanning electron microscope (SEM) image of the $Bi_2Se_3$ nanowire mechanical resonator. Electrical conductance is measured by comparing drain current ($I_D$) and source-drain voltage ($V_{SD}$). DC and radio-frequency (RF) voltages ($V_g$ and $V_{RF}$ respectively) are applied to the gate electrode, with RF voltage at the drain electrode amplified and recorded ($V_{OUT}$) to actuate and detect mechanical resonance. A magnetic field ($B$) is applied parallel to the nanowire axis to study AB oscillation due to TI surface states. **b**, The equivalent circuit for RF frequency involves mechanically compliant geometric capacitance ($C_G$) and quantum capacitance ($C_Q$) of the nanowire. **c**, Colourmap of $V_{OUT}$ for a range of $V_g$. The parabolic shift in resonance is due to the effective spring constant change dominated by $C_G$. Inset:

Example of mechanical resonance with a resonant frequency of 115.09 MHz and quality factor of $1.2 \times 10^4$ at $V_g = -28.0$ V. The solid red line is a Lorentzian fit. **d**, Energy dispersion of surface states in a TI nanowire when magnetic flux $\Phi$ is zero or $\Phi_0/2$ where $\Phi_0$ is the flux quantum ($= h/e$). Red lines indicate the non-degenerate gapless mode. **e**, Expected conductance ($G$) and density of states (DOS) of the surface states as a function of chemical potential $\mu$ in units of 1D sub-band gap $\Delta$, when the magnetic flux is zero (black) or $\Phi_0/2$ (red). **f**, Mechanical resonant frequency shifts ($\Delta f_0$) as a result of DOS oscillation via quantum capacitance effects. The resonant frequency shift is calculated from equation (5) considering DOS oscillation in (**e**), and its two terms $\Delta f_\mathrm{I}$ and $\Delta f_\mathrm{II}$ are proportional to $Q^2$ and $Q^3$, respectively.

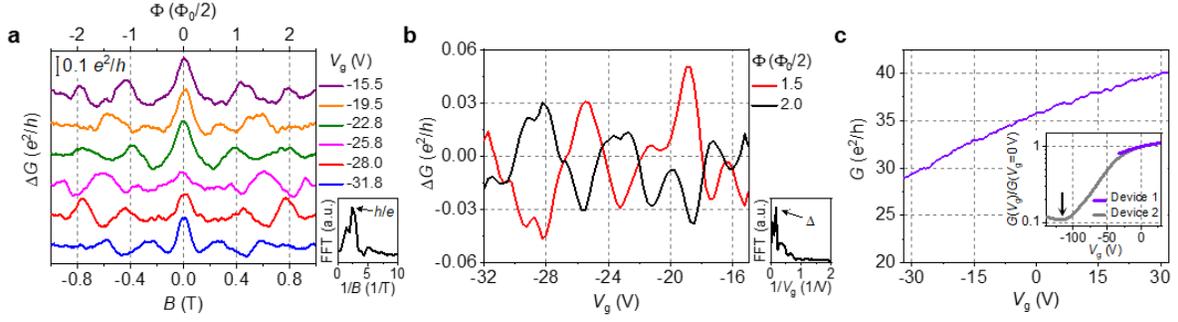

**Figure 2| Conductance oscillation as a function of magnetic field and gate voltage. a**, Conductance oscillations as a function of magnetic field at different gate voltages after subtracting background conductance. The $\Delta G$ oscillations have a period of $\Delta B = 0.4$ T as identified from the first peak in the FFT spectrum (inset). $\Delta B$ corresponds to one flux quantum ($= h/e$) across the nanowire cross-section, confirming AB oscillation. The magneto-conductance oscillations show periodic alternations with respect to gate voltages, as expected from the Berry phase in the surface-state Hamiltonian (equation (2)). Inset: FFT spectrum of $\Delta G$ vs. $B$ at $V_g = -22.8$ V. Arrows indicate the peak position at period $\Delta B = 0.4$ T. **b**, $\Delta G$ as a function of gate voltage $V_g$ at two representative magnetic fluxes. The red line is the half-integer flux quantum ($B = 0.6$ T) and the black line is the integer flux quantum ($B = 0.8$ T). The period of the $\Delta G$ oscillations with respect to gate voltage is $\Delta V_g \approx 5.7$ V. Inset: FFT spectrum of $\Delta G$ vs. $V_g$ at $B = 0.6$ T. The peak is located at period $\Delta V_g = 5.7$ V. **c**, Total conductance $G$ vs. gate voltage at zero magnetic field. Inset: By comparing to another device with identical geometry (Device 2), the Dirac point is expected to be at $V_{g0} \approx -113$V.

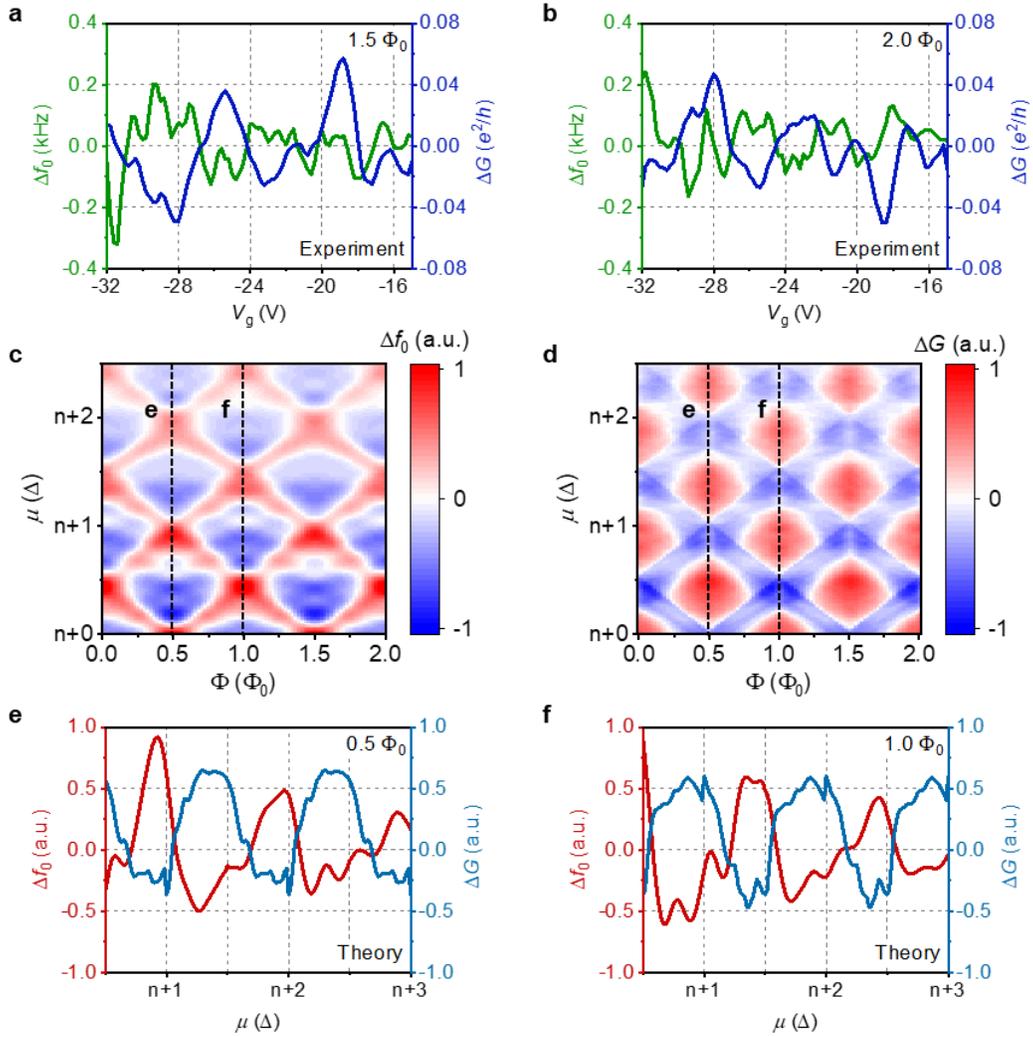

**Figure 3| AB oscillations in the mechanical resonant frequency and conductance of a TI nanowire. a, b**, Measured mechanical resonant frequency shift $\Delta f_0$ and conductance modulation $\Delta G$ as a function of gate voltage $V_g$ at half-integer flux quanta and integer flux quanta, showing an out-of-phase relation. **c, d**, Model calculations of $\Delta f_0$ and $\Delta G$ plotted as a function of chemical potential $\mu$ and magnetic flux $\Phi$. **e, f**, Calculated $\Delta f_0$ and $\Delta G$ at half-integer and integer flux quanta conditions demonstrating an out-of-phase relation, plotted from the corresponding vertical cuts in (**c**) and (**d**).

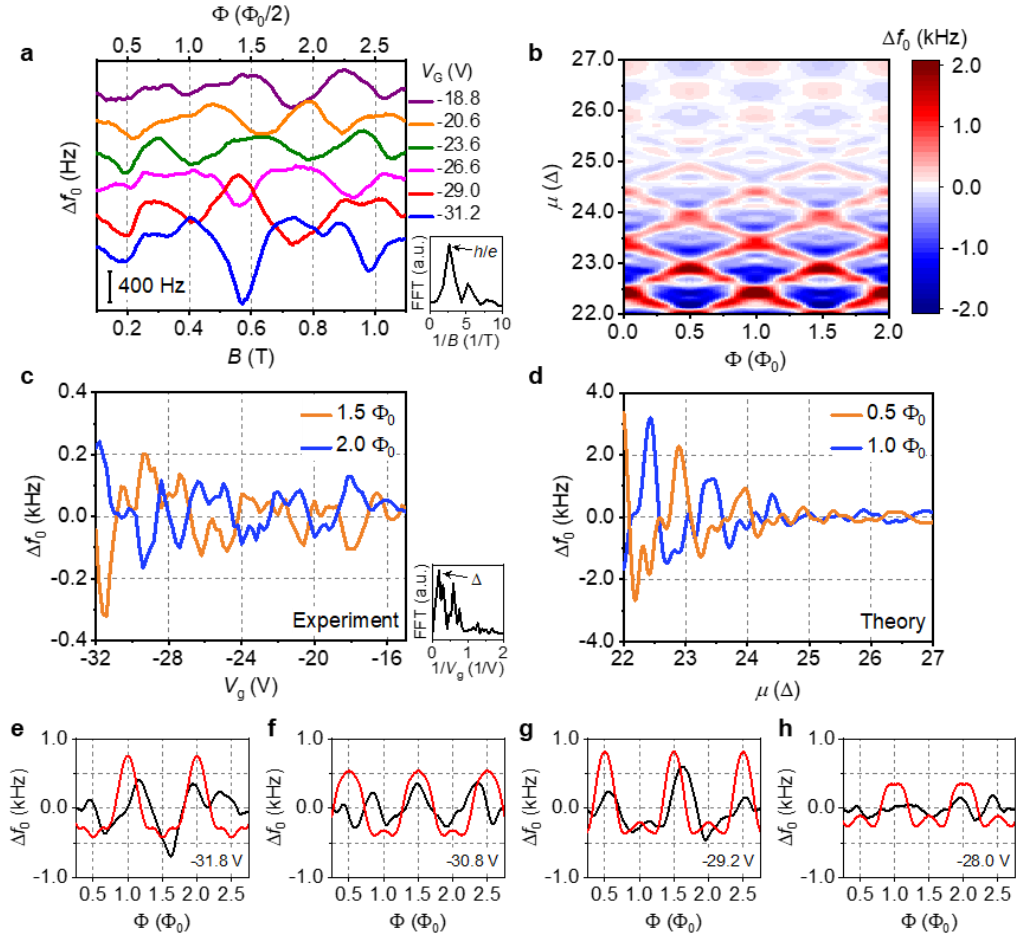

**Figure 4| AB oscillations in the mechanical resonant frequency of a TI nanowire. a**, Measured mechanical resonance shift $\Delta f_0$ as a function of magnetic field at different gate voltages. The $\Delta f_0$ oscillates with a period of $\Delta B = 0.4$ T as identified from the first peak in the FFT spectrum of $\Delta f_0$ vs. $B$ (inset), identical to the period observed in conductance measurements. Inset: FFT spectrum of $\Delta f_0$ vs. $B$ at $V_g = -26.6$ V. Arrows indicate the peak position at period $\Delta B = 0.4$ T. **b**, Model calculations of mechanical resonant frequency shift plotted as a function of chemical potential $\mu$ and magnetic flux $\Phi$ showing a change in oscillation pattern. **c, d**, Measured (**c**) and calculated (**d**) resonant frequency shift as a function of gate voltages $V_g$ and chemical potential $\mu$ respectively, at half-integer (orange) and integer (blue) flux quanta. Inset in **c**: FFT spectrum of $\Delta f_0$ vs. $V_g$ at $B =$

0.6 T. The peak is located at period $\Delta V_g$ = 5.7 V, identical to the period observed in conductance measurements. **e–h**, Calculated (red) and measured (black) resonant frequency shifts as a function of magnetic flux at gate voltages $V_g$ = –31.8 V (**e**), –30.8 V (**f**), –29.2 V (**g**), and –28.0 V (**h**). Corresponding chemical potentials are 23.53, 23.82, 23.91, and 24.33, respectively, in the units of $\Delta$.

# Supplementary Information

**Nanomechanical characterization of quantum interference in a topological insulator nanowire**

**Authors:** Minjin Kim[1], Jihwan Kim[2], Yasen Hou[3], Dong Yu[3], Yong-Joo Doh[4], Bongsoo Kim[1], Kun Woo Kim[5*] and Junho Suh[2*]

**Affiliations:**

[1]Department of Chemistry, Korea Advanced Institute of Science and Technology, Korea

[2]Quantum Technology Institute, Korea Research Institute of Standards and Science, Korea

[3]Department of Physics, University of California at Davis, U.S.A.

[4]Department of Physics and Photon Science, Gwangju Institute of Science and Technology, Korea

[5]Center for Theoretical Physics of Complex Systems, Institute for Basic Science (IBS), Daejeon, Korea

*Correspondence to: kkimx4@ibs.re.kr, junho.suh@kriss.re.kr

Contents

# S1. Characterization of the Bi$_2$Se$_3$ nanowire

## S1.1. Bi$_2$Se$_3$ nanowire dimensions

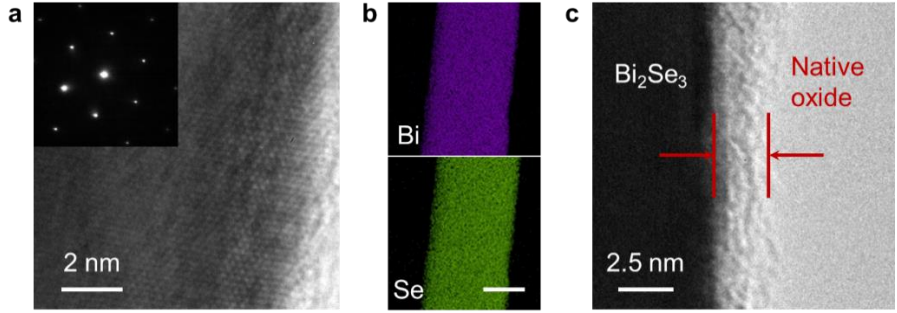

**Figure S1. a,** High-resolution transmission electron microscope (TEM) image and (inset) selected area electron diffraction pattern of the single-crystalline Bi$_2$Se$_3$ nanowire. **b,** Energy-dispersive X-ray spectroscopy elemental maps showing homogeneous atomic distribution of Bi (violet) and Se (green). Scale bar, 100 nm. **c,** TEM image of the Bi$_2$Se$_3$ nanowire indicating a 2.5-nm-thick native oxide layer is present on its surface. Cross-sectional area $S$ is calculated by $S$ = (width – 2×native oxide layer thickness) × (thickness – 2×native oxide layer thickness). With a native oxide thickness of 2.5 nm, the nanowire in Device 1 has dimensions of width = 105 nm and thickness = 116 nm with $S = 1.11\times10^{-14}$ m$^2$. The $\Delta B$ deduced from the cross-sectional area $S$ of Device 1 is 0.37 T, which is given by $\Delta B = \Phi_0/S$. The effective cross-section is $\Phi_0/\Delta B = 1.04\times10^4$ nm$^2$, which is slightly smaller than the actual value possibly due to the finite penetration depth of boundary states.

## S1.2 Gate response and Dirac point

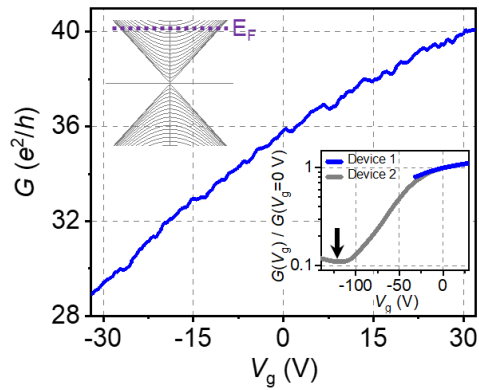

**Figure S2.** Conductance vs. gate voltage in Device 1. Inset: From comparison with another device of identical geometry (Device 2), the Dirac point is estimated to be at $V_g \approx -113$ V.

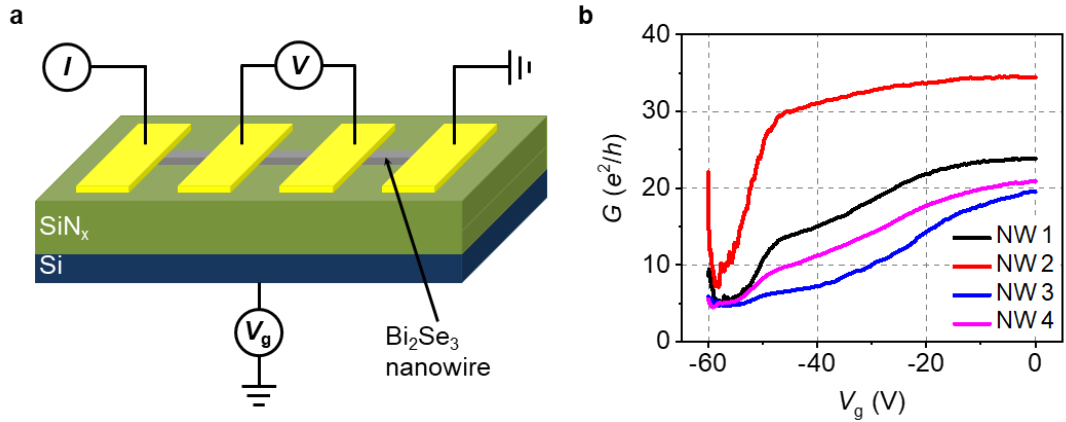

**Figure S3. a,** Conductance measurement schematic showing the $Bi_2Se_3$ nanowire on a $SiN_x$/Si substrate with back-gate voltage $V_g$. **b,** Nanowire conductance $G$ response to $V_g$. The four different nanowires exhibit Dirac points at $V_g = -57.6$ V (NW 1), $-58.1$ V (NW 2), $-57.0$ V (NW 3), and $-59.1$ V (NW 4).

In Fig. S2, the nanowire conductance of Device 1 decreases with decreasing gate voltage, showing n-type semiconductor behavior. Because Device 1 was destroyed during gating, its Dirac point ($V_g = -113$ V) was estimated from another similarly fabricated nanowire (Device 2). Even though two nanowires may have a different circumference and impurity concentration, their intrinsic chemical potential $\mu_0$ would be close to each other as it is determined by average chemical composition. As such, the gate voltage required to bring the chemical potential all the way down to the Dirac point is similar for each nanowire, because the DOS per unit length and the induced charge are both linearly proportional to nanowire circumference. This argument is supported by additional measurements of four more nanowires that are in contact with the substrate instead of being suspended (Fig. S3a). The gate voltage at the Dirac point is different ($V_g \sim -58$ V) on account of the different geometry, but it is consistent among all four nanowires (Fig. S3b). It is interesting to note that the wires have different total conductances even though their intrinsic chemical potentials are supposedly the same. This might be due to differences in device-specific impurity concentration; while the amount of induced charge (integration of the DOS) is less sensitive to impurities, the total conductance could be significantly reduced (see S6 for conductance calculations for different disorder strengths).

## S2. Mechanical resonant frequency shift measurement

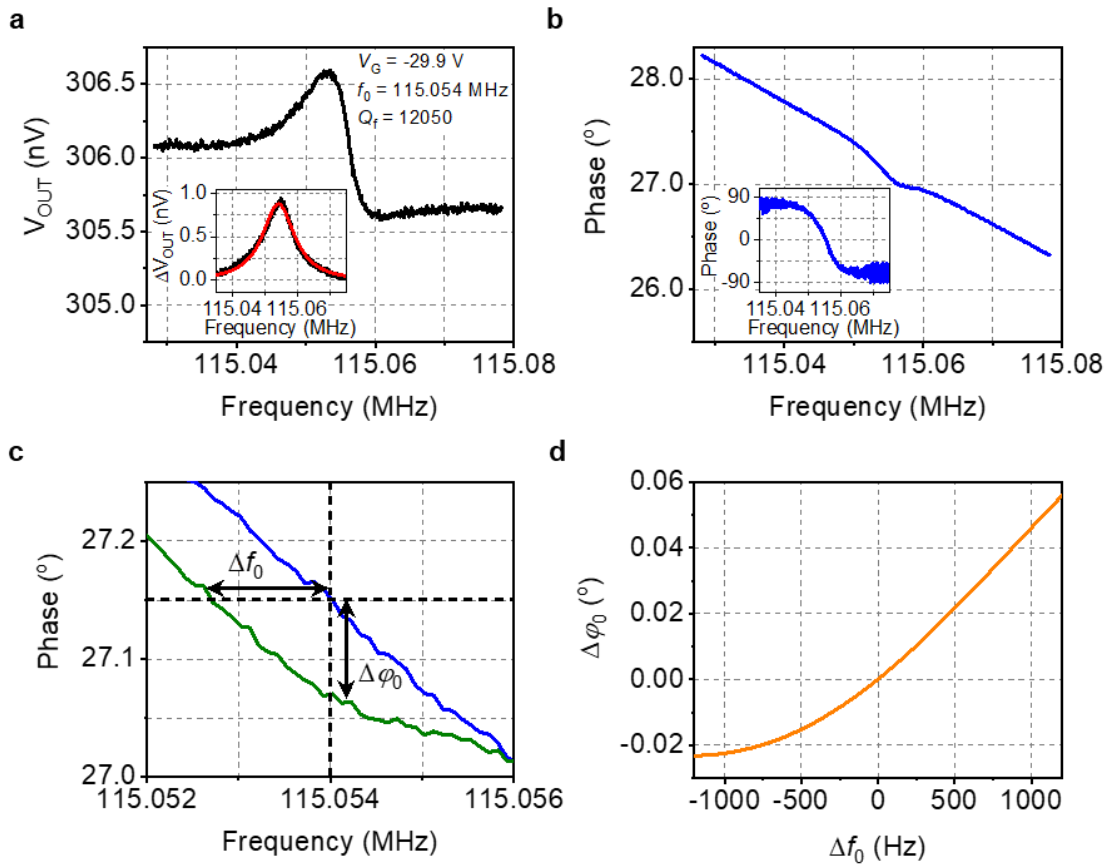

**Figure S4. a,** Amplitude response of mechanical resonance. Inset: Amplitude response after subtracting background response. The solid red line is a Lorentzian fit with resonance frequency $f_0$ and quality factor $Q_f$. **b,** Phase response of mechanical resonance. Inset: Phase response after subtracting background response. **c,** Phase response change (e.g. from blue to green) when $f_0$ is modulated by $\Delta f_0$ ($\ll f_0/Q_f$), and the phase at $f_0$ shifts by $\Delta\varphi_0$. **d,** Extracted relation between $\Delta f_0$ and $\Delta\varphi_0$.

From our device, we obtain electric signal $V_{OUT}$ oscillating at the frequency of applied radio-frequency voltage $V_{RF}$, but with a phase difference. The amplitude and phase of $V_{OUT}$ is plotted in Fig. S4a–b, respectively. A typical mechanical resonant frequency $f_0$ is approximately 115 MHz and quality factor $Q_f$ is $\sim 1.2 \times 10^4$. After subtracting the background, a Lorentzian function of amplitude near the resonance and a 180° change of phase are plotted in the insets. In Fig. S4c, the resonant frequency shift induced by a threading magnetic field and corresponding phase shift at the resonant frequency are shown. Measurements of the phase shifts for a given $V_{RF}$ and $V_g$ were made by varying the magnetic field. The frequency shift is obtained based on the relation plotted in Fig. S4d.

## S3. Modeling and calculation of geometric capacitance

Multiple experimental observations provided guidance for our numerical modeling: (i) an approximated chemical potential ($\sim 24\,\Delta$) estimated from the measurement of total conductance, Fig. 2c; (ii) the periodicity ($\sim 5.7$ V/$\Delta$) of Aharonov–Bohm (AB) oscillation along the energy, Fig. 2b; (iii) the value of $\ddot{C}_G$ at $V_g = 0$ obtained from the quadratic functional shape of the resonant frequency shift, Fig. 1c; and (iv) the out-of-phase relation between magneto-conductance and resonant frequency shift, Fig. 3a–b, which indicates $\Delta f_I < \Delta f_{II}$. In this section, we devise a geometric capacitance model that satisfies the above conditions.

### S3.1. Geometric capacitance

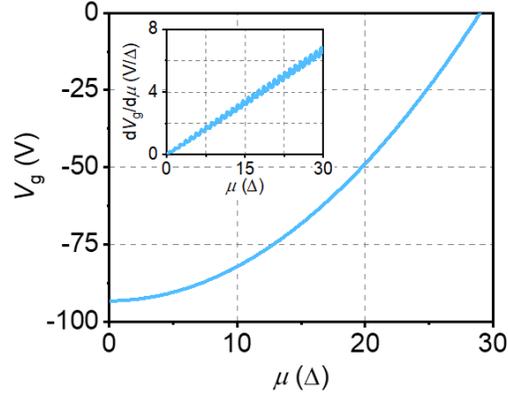

**Figure S5.** Calculated relation of gate voltage $V_g$ and chemical potential $\mu$ based on geometric capacitance $C_G = 1.5 \times 10^{-17}$ F. Inset: $dV_g/d\mu$ value of about 5.7 V/$\Delta$ at a $\mu$ range of 23 $\Delta$ to 25 $\Delta$. This chemical potential ($\mu$) range corresponds to a $V_g$ range of –32 V to –15 V.

Figure S5 plots the relation between gate voltage and chemical potential ($\mu_0 = 29\,\Delta$ is chosen) according to Kirchhoff's circuit law based on a geometric capacitance of $C_G = 1.5 \times 10^{-17}$ F. The oscillation of $dV_g/d\mu$ (inset) reflects the modulation of the DOS, which is computed with energy broadening $\eta = 0.2\,\Delta$ (see S5 for details). Near $V_g = -30$ (V), we can see that chemical potential is $\mu \sim 24\,\Delta$ and $\frac{dV_g}{d\mu} \sim 5.5$ (V/$\Delta$), consistent with experimental values.

Our device setup contains two metals separate in space, namely the $Bi_2Se_3$ nanowire and the gate plate, so geometric capacitance is the ratio of induced charge and voltage difference between them. We adopt the following two-parallel-plate model:

$$C_G(x) = \frac{\varepsilon_0 A}{d + \alpha x},$$

where $A$ is effective area, $d$ is distance, $x$ is displacement during the vibration, and $\alpha$ (= 0.523) is a geometrical correction as neither end of the nanowire moves[1]. From this, derivatives are necessary to compute the resonant frequency shift (see S5 for details):

$$\dot{C}_G = -\alpha \frac{\varepsilon_0 A}{(d+\alpha x)^2}, \qquad \ddot{C}_G = 2\alpha^2 \frac{\varepsilon_0 A}{(d+\alpha x)^3}.$$

As seen in Fig. 1c, the shift of resonant frequency is nearly symmetric with $V_g$. This overall shape is deduced from the classical circuit model where the change of chemical potential is ignored ($\mu = \mu_0$):

$$\delta U_{ec} = \delta\left(\frac{1}{2} C_G V_g^2\right) - V_g \delta Q = -\delta\left(\frac{1}{2} C_G V_g^2\right).$$

Here, the first term is the charging energy of the capacitor, and the second term is work done to the battery for a fixed gate voltage, $\delta Q = V_g \delta C_G$. Thus, the change in spring constant is $\Delta k = \ddot{U}_{ec} = -\frac{1}{2}\ddot{C}_G V_g^2$, and a parabolic shape with respect to $V_g$ naturally appears. From our measurement of resonant frequency shift:

$$\Delta f = \frac{f_0}{2k_0} \Delta k = -\frac{f_0}{4k_0} \ddot{C}_G V_g^2,$$

where $k_0 = (2\pi f_0)^2 m_{eff}$ and $m_{eff} = \left(\frac{0.397}{0.523^2}\right) m$, where $m = 1.246 \times 10^{-16}$ kg and the numerical coefficient is again due to the way the nanowire vibrates in its fundamental flexural mode[1]. $\ddot{C}_G = 9.91 \times 10^{-4}$ F/m$^2$ is estimated from Fig. 1c. Following the two-parallel-plate model, we use $d = 174$ nm. Since vibration amplitude is much smaller than distance as estimated in S3.3, $x \ll d$, we set $x = 0$. As a result, $\dot{C}_G = -1.65 \times 10^{-10}$ F/m. On the other hand, we found $C_G = 1.5 \times 10^{-17}$ F is needed to explain 5.7 V/Δ, while the value obtained from integration is $C'_G = \ddot{C}_G \frac{d^2}{2\alpha^2} = 5.48 \times 10^{-17}$ F. Such difference in geometric capacitance may come from an additional capacitance, $C_0$, which is uninvolved in the mechanical vibration but connected in series to the TI nanowire. Together, $C_G = C'_G C_0 / (C'_G + C_0)$ yields the desired geometric capacitance with $C_0 = 2.07 \times 10^{-17}$ F.

## S3.2. Displacement of nanowire equilibrium position by gate voltage

With the application of gate voltage, the distance between the nanowire and gate would change by the force associated with the electrostatic energy by

$$F = -\dot{U}_{ec} - k_0 \Delta x_0 = \frac{1}{2}\dot{C}_G V_g^2 - k_0 \Delta x_0,$$

which is zero at equilibrium position. We get $\Delta x_0 = 0.89$ nm. This displacement is much smaller than the nanowire length (1.5 μm) and the distance between the nanowire and gate electrode ($d \sim$ 100 nm). Thus, we assume that the magnetic flux direction is parallel to the current direction along the nanowire, and in the calculation of the resonant frequency shift, we neglect the change of equilibrium position.

## S3.3. Amplitude of mechanical vibration of the nanowire

A driven mechanical oscillator with natural frequency may have a large oscillating amplitude:

$$x_{RF} = \frac{F_{RF}/m_{eff}}{\sqrt{(\omega_0^2 - \omega^2)^2 + \omega^2\beta^2}},$$

where force induced by the voltage source with radio frequency is $F_{RF} \cong \dot{C}_G V_g V_{RF}$ and the Q-factor is defined by $Q_f = \sqrt{\omega_0^2 - (\beta^2/2)}/\beta \cong \omega_0/\beta$ for a system with a high quality factor (damping rate $\beta$ is much smaller than $\omega_0$). From Fig. 1c, our device has $Q_f \cong 11900$. We apply $V_{RF} = 0.4$ mV, which is much smaller than the gate voltage ($V_g \sim -30$ V) to maintain the small mechanical vibration amplitude $x_{RF} \approx 265$ pm, which is to minimize chemical potential change arising during vibration.

## S4. Density-of-states calculation from the continuum model

Numerical calculation of the DOS of a quasi-1D wire is explained in the *Methods* section of the main text. Here, we add a few more DOS maps to provide more insight.

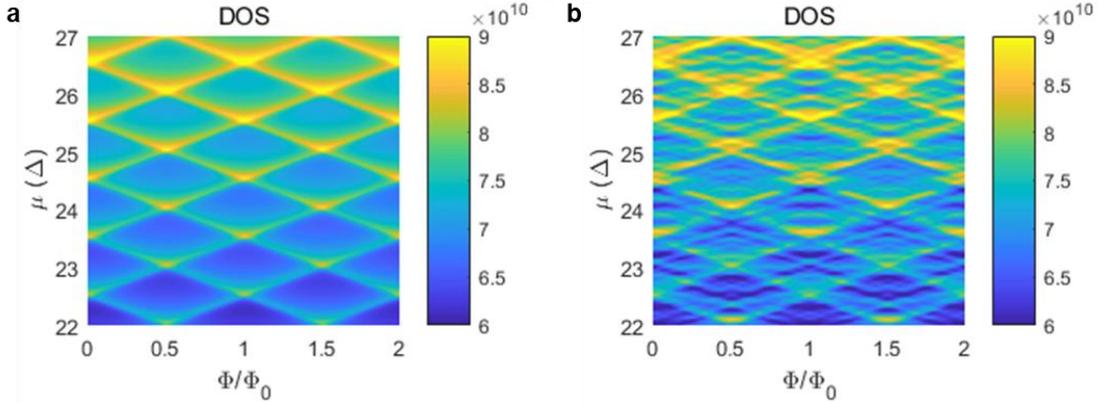

**Figure S6.** DOS (the number of states per unit length per unit energy eV) of a quasi-1D wire for $\eta = 0.05\,\Delta$ when (**a**) wire length is much longer than circumference, $L/(2\pi R) \gg 1$ and (**b**) wire length is three times the circumference, $L/(2\pi R) \cong 3$.

For the energy broadening $\eta = 0.05\,\Delta$ used in Fig. 1e, DOS maps are generated in the domain of magnetic flux and chemical potential. When wire length is much longer than circumference, $L/(2\pi R) \gg 1$ (Fig. S6a), the DOS shows peaks whenever a new transverse mode is encountered, as the band bottom of a quasi-1D mode has a diverging DOS ($\sim 1/v_F$). The Fig. S6b figure reflects the DOS of the actual dimension ratio, $L/(2\pi R) \cong 3$. On top of the diamond-shaped lines, there are additional modulations of the DOS due to the discretization of eigenmodes along the direction of current propagation, $\Delta' \cong \hbar v_F/L = \Delta/3$. While conductance is dependent on the number of transverse modes, mechanical resonance shift is influenced by all details of the DOS. While this may complicate characterization of the nanowire, it also provides more information on microscopic details.

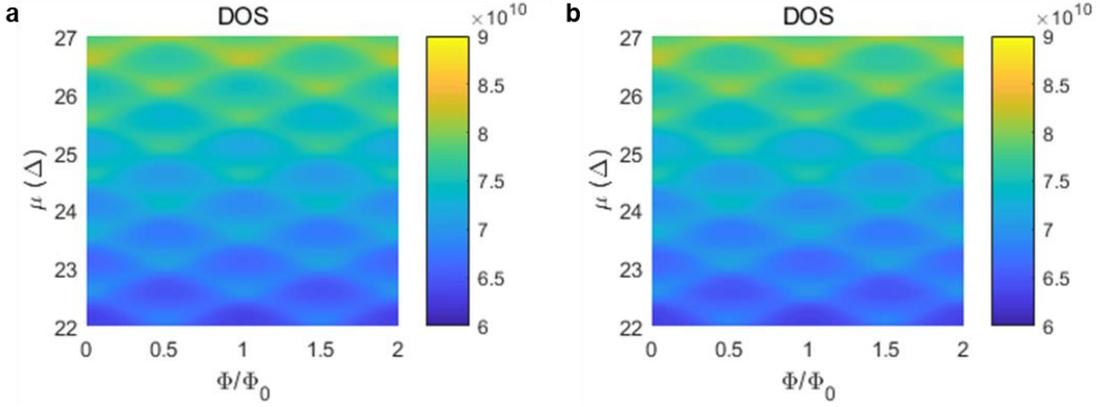

**Figure S7.** DOS (the number of states per unit length per unit energy eV) of of a quasi-1D wire for $\eta = 0.20\,\Delta$ when (**a**) wire length is much longer than circumference, $L/(2\pi R) \gg 1$ and (**b**) wire length is three times the circumference, $L/(2\pi R) \cong 3$.

We present DOS maps for $\eta = 0.20\,\Delta$ (Fig. S7) for the two geometries used for calculation in the main text. Here, modulation of the DOS is smoothed out, and detailed features of the eigenmodes for the $L/(2\pi R) = 3$ case are significantly reduced. The quantum capacitance is

$$C_Q = Le^2 \nu(E) \cong (1.5 \times 10^{-6}) \times (1.6 \times 10^{-19})^2 \left(\frac{50}{\hbar v_\text{F}}\right) = 3.6 \times 10^{-14}\,\text{F},$$

where $\nu(\mu \cong 25\Delta)$ is approximated to the DOS of fifty 1D wire with Fermi velocity[2] $v_\text{F} = 5\times 10^5$ m/s. As claimed in the manuscript, in the energy window of our measurement the quantum capacitance is ~1000 times larger than the geometric capacitance.

## S5. Calculation of mechanical resonant frequency shift

To compute the shift in the mechanical resonant frequency of a TI nanowire, we need to know the change in total energy caused by the flexural mode:

$$\delta U_\text{tot} = \delta\left(\frac{1}{2} m\dot{x}^2 + \frac{1}{2} k_0 x^2\right) + \delta U_\text{ec},$$

$$\delta U_{ec} = \delta\left(\frac{Q^2}{2C_G}\right) - V_g \delta Q + \delta\left(Q\frac{\mu - \mu_0}{e}\right),$$

where in the first line, total energy is the sum of mechanical energy and electrostatic energy $\delta U_{ec}$, and in the second line, $\delta U_{ec}$ contains three circuit elements: charging energy of the geometrical capacitor, work done to the gate battery, and charging energy of the TI nanowire. Note that the last term cannot be replaced by $\delta(Q^2/2C_Q)$ as quantum capacitance is a nonlinear function of induced charge. Since our measurements are made by varying gate voltage and magnetic field, hereafter we focus on the resonant frequency shift induced by changes in electrostatic energy.

Our goal is to simply compute the second derivative of $U_{ec}$ with respect to $x$. Several quantities are dependent on $x$. First, as discussed in S3, geometric capacitance is position-dependent. For a fixed gate voltage, induced charge $Q$ is position-dependent and so is the resulting change of chemical potential $\mu$. Two relations are helpful to simplify further calculation; one is Kirchhoff's law as shown in the main text, and the other is its derivative for a fixed gate voltage:

$$\frac{\partial Q}{\partial x} = \frac{C_Q}{C_G}\frac{Q}{C_G + C_Q}\frac{\partial C_G}{\partial x},$$

where change in chemical potential is expressed in terms of induced charge and quantum capacitance (the DOS) by:

$$\frac{1}{e}\frac{\partial \mu}{\partial x} = \frac{1}{C_Q}\frac{\partial Q}{\partial x},$$

where $C_Q = e^2 L\nu$ and $\nu = \delta(Q/Le)/\delta\mu$. Electrostatic energy is first simplified by Kirchhoff's law:

$$\frac{dU_{ec}}{dx} = -\frac{Q^2}{2C_G^2}\frac{\partial C_G}{\partial x} + \frac{Q}{C_Q}\frac{\partial Q}{\partial x},$$

$$= -\frac{Q^2}{2C_G^2}\frac{\partial C_G}{\partial x} + \frac{1}{C_G}\frac{Q^2}{C_G + C_Q}\frac{\partial C_G}{\partial x}.$$

By taking one more derivative, we obtain the desired result:

$$\frac{d^2 U_{ec}}{dx^2} = \frac{1}{C_G^2}\left[\frac{2C_Q - C_G}{(C_G + C_Q)^2}\dot{C}_G^{\,2} + \frac{C_G - C_Q}{2(C_G + C_Q)}\ddot{C}_G\right]Q^2 + \frac{1}{2}\left(\frac{\dot{C}_G}{C_G}\right)^2 e\frac{\partial}{\partial \mu}\left(\frac{1}{(C_G + C_Q)^2}\right)Q^3,$$

$$= k_{\mathrm{I}} + k_{\mathrm{II}},$$

where the dots indicate the derivatives with respect to displacement *x*. So far, no approximation has been made except for the electrostatic condition of our circuit, assuming that mechanical vibration is much slower than the time required for electrical equilibration. The above expression is used for Fig. 4. Since our device has $C_G \ll C_Q$ and $\dot{C}_G^{\,2} \cong C_G \ddot{C}_G$, the expression is simplified to the one shown in the main text.

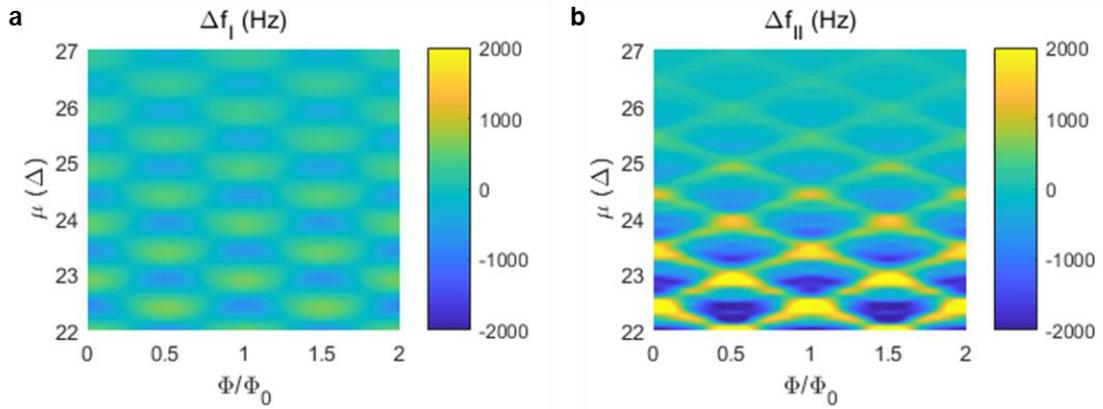

**Figure S8.** Mechanical resonant frequency shifts ($\Delta f_0$) as a result of DOS oscillation via quantum capacitance effects.

In Fig. S8, two contributions of resonant frequency shift $\Delta f_{\mathrm{I,II}}$ are plotted separately to show their qualitative difference in modulation and scaling over energy.

## S6. Magneto-conductance for different disorder strengths

The *Methods* section in the main text contains details of conductance calculation. Here, we add conductance maps for different disorder strengths.

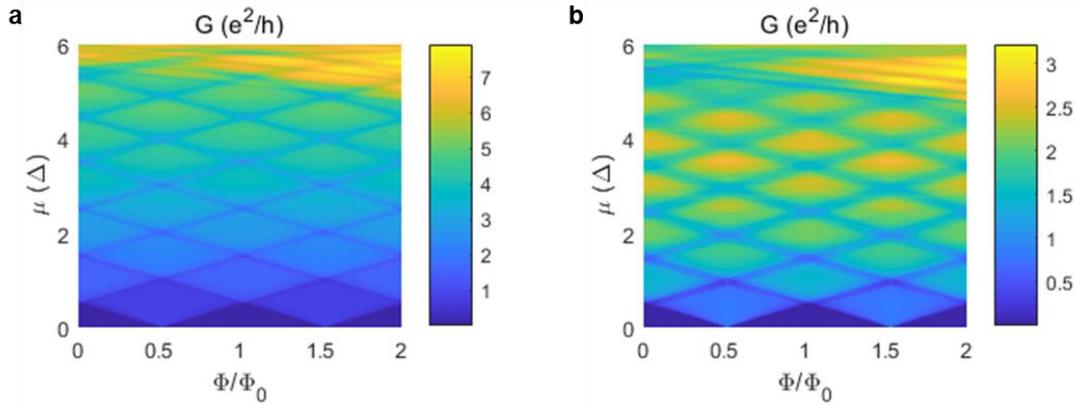

**Figure S9.** Conductance maps at disorder strengths of (**a**) $W = 1.5\,\Delta$ and (**b**) $W = 3.0\,\Delta$.

Without disorder, the band gap is at $\mu = 6.0\,\Delta$, where $\Delta$ is the level spacing between neighboring transverse modes at $k = 0$. The Fig. S9 is conductance maps at disorder strengths of $W = 1.5\,\Delta$ and $3.0\,\Delta$. The horizontal axis is magnetic flux from zero to two quanta. The AB modulation of conductance is clearly seen along the energy (vertical axis), as well as along the magnetic flux. As opposed to a step-like jump of conductance in a clean nanowire, in disordered cases we can see dips develop between conductance plateaus. This indicates that, in the presence of onsite random chemical potential, and near the energy following the diamond-shaped lines, the number of transverse modes engaged in transport fluctuates, thereby causing backscattering. The width of such dips widens for $W = 3.0\,\Delta$.

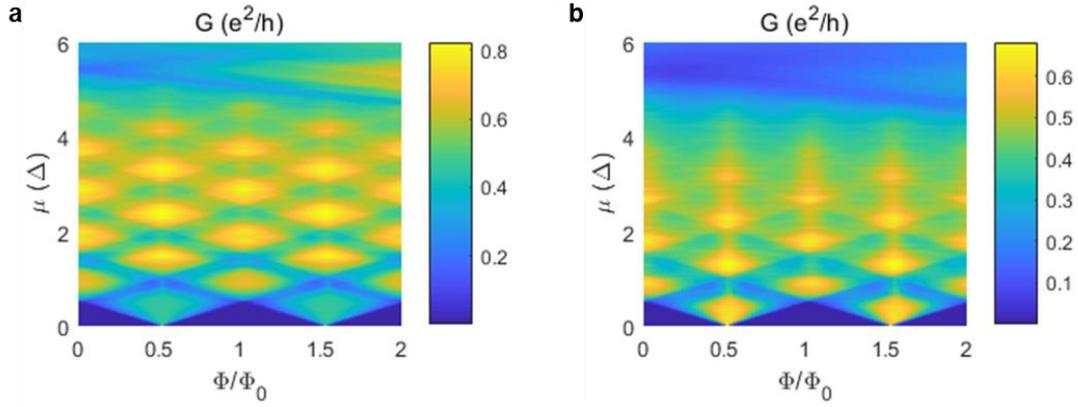

**Figure S10.** Conductance maps at disorder strengths of (**a**) $W = 4.5\,\Delta$ and (**b**) $W = 6.0\,\Delta$.

Next, the Fig. S10 is conductance maps for disorder strengths of $W = 4.5\,\Delta$ and $6.0\,\Delta$. Quite remarkably, a feature of weak anti-localization or so-called Altschuler–Aronov–Spivak (AAS) oscillation[3] begins to develop as the crossover from a quasi-1D wire to diffusive 2D transport takes place, when the energy broadening by disorder is larger than the energy spacing between neighboring transverse modes. The period of AAS oscillation along the magnetic flux is $\Phi = \pi$, and it is energy independent. The right figure above shows clear 'pillars' of conductance along the energy.

For Gaussian-correlated impurities in continuum space, $\langle V(r')V(r)\rangle = K_0 \frac{(\hbar v_F)^2}{2\pi\xi^2} e^{-|r-r'|^2/2\xi^2}$, the elastic linewidth broadening $\eta$ is given by[4]:

$$\frac{\hbar}{\tau} = \int \frac{dq\,dr}{4\pi}\left(1 - \cos^2\theta_q\right)\delta(k_F - q)\langle V(0)V(r)\rangle e^{iq\cdot r},$$

$$= \frac{\hbar v_F}{\xi}\frac{2K_0 I_1[(k_F\xi)^2]}{k_F\xi \exp[(k_F\xi)^2]},$$

where $k_F$ is a wavenumber at Femi energy, $K_0 = \frac{1}{3}\gamma_0^2/(\hbar v_F d)^2$ is dimensionless disorder strength, $d$ is mean distance between impurities, and $I_1$ is a modified Bessel function. We want to compute this quantity for our lattice model. We set onsite random chemical potential on every lattice with uniform distribution $V_n \in [-W, W]$, $H_{dis} = \sum_{n \in S} \psi_n^\dagger (V_n \Gamma^0) \psi_n$ instead of impurities with Gaussian profiles spread in space. In this case, with the lattice spacing $a$, $d = a$, $W = \gamma_0 a^2$, $\xi = a$ (spatial resolution), $\Delta = \frac{\hbar v_F}{R}$ (transverse mode energy-level spacing), and $k_F \xi = \frac{E}{\Delta} \frac{a}{R}$. Collecting all,

$$\frac{\hbar}{\tau}\left(\frac{1}{\Delta}\right) = \frac{2}{3}\left(\frac{W}{\Delta}\right)^2 \left(\frac{\Delta}{E}\right) \frac{I_1[(k_F \xi)^2]}{exp[(k_F \xi)^2]}.$$

For example, when $\frac{\hbar}{\tau} = 0.2\,\Delta$, the scattering time is $\tau = 0.7$ ps. Note that the scattering time goes to infinity at the Dirac point as $I_1(x) \sim x$ for $x \ll 1$. Linewidth broadening $\eta$ for four disorder strengths is plotted Fig. S11.

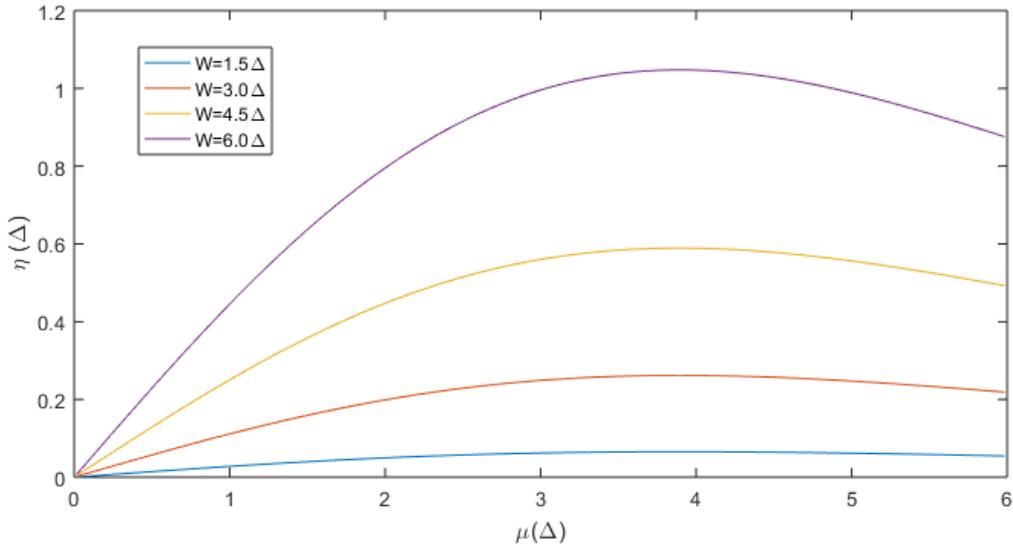

**Figure S11.** Linewidth broadening $\eta$ for four disorder strengths of $W = 1.5\,\Delta$, $3.0\,\Delta$, $4.5\,\Delta$, and $6.0\,\Delta$.

For $W = 6.0\,\Delta$, as the energy broadening becomes comparable to the energy level spacing of neighboring transverse modes, the crossover from quasi-1D to diffusive 2D transport can be observed, with our conductance map (Fig. S10b) clearly showing AAS oscillation as well as AB oscillation with increasing Fermi energy. Lastly, we comment that the energy broadening factor ($\eta = 0.20\,\Delta$) used in our DOS calculation corresponds to an impurity strength between $W = 3.0\,\Delta$ and $W = 4.5\,\Delta$. Even though the resonant frequency shift map and the conductance map in Fig. 3 are obtained in different energy windows, the energy broadening used for the former is consistent with the disorder strength used for the latter.